\documentclass[acmtocl]{acmsmall}

\usepackage{cdgcommands}

\usepackage{color}

\long\def\comment#1{}

\title{On the expressive power of multiple heads in CHR}

\author{CINZIA DI GIUSTO
\affil{INRIA Rh\^{o}ne Alpes, Grenoble, France}
MAURIZIO GABBRIELLI
\affil{Dipartimento di Scienze dell'Informazione and Lab. Focus INRIA, Universit\`a di Bologna, Italy}
MARIA CHIARA MEO 
\affil{Dipartimento di Scienze, Universit\`a di Chieti Pescara, Italy}}

\begin{abstract}
Constraint Handling Rules (CHR) is a committed-choice declarative
language which has been originally designed for writing constraint solvers and which is nowadays a general purpose language.
CHR programs consist of multi-headed guarded rules which allow
to rewrite constraints into simpler ones until a solved form is
reached. Many empirical evidences suggest that multiple heads augment the expressive power of the language, however no formal result in this direction has been proved, so far.

In the first part of this paper we analyze the Turing completeness of CHR with respect to the underlying constraint theory. We prove that if the constraint theory is powerful enough then restricting to single head rules does not affect the Turing
completeness of the language. On the other hand, differently from the case of the multi-headed language, the single head CHR language is not Turing powerful when the underlying signature (for the constraint theory) does not contain function symbols.

In the second part we prove that, no matter which constraint theory is considered,
under some reasonable assumptions it is not possible to encode the CHR language (with multi-headed rules) into a single headed language while preserving the semantics of the programs. We also show that, under some stronger assumptions,
considering an increasing number of atoms in the head of a rule augments the expressive power of the language.
These results provide a formal proof for the claim that multiple heads augment the expressive power of the CHR language.
\end{abstract}




\begin{document}

\begin{bottomstuff}
The research of Cinzia Di Giusto is partially supported by Fondation de Coop\'eration Scientifique Digiteo Triangle de la Physique. 
The research of Maurizio Gabbrielli and Maria Chiara Meo has been partially supported by the PRIN project  20089M932N,
``Innovative and multi-disciplinary approaches for constraint and preference reasoning''. 
\end{bottomstuff}

\maketitle

\section{Introduction}
	Constraint Handling Rules (CHR) \cite{Fr91,Fruhwirth98} is a
com\-mit\-ted-choice declarative language which has been originally designed for writing constraint solvers and which is nowadays a general purpose language. A CHR program consists of a set of multi-headed guar\-ded
(simplification and propagation) rules which allow
to rewrite constraints into simpler ones until a solved form is
reached. The language is parametric with respect to an underlying constraint theory CT which defines the meaning of basic built-in constraints.

% are speci\-fically
%designed to implement the two most important
%operations involved
%in constraint solving processes.
%Simplification rules allow to
%replace constraints by simpler ones, while preserving their
%meaning. Propagation rules are used to add new redundant
%constraints which do not modify the meaning of the given
%constraint but that can be useful for further reductions.

The presence of multiple heads  is a crucial feature  which differentiates CHR from
other existing committed choice (logic) languages.
Many examples in the vast literature on CHR provide empirical evidence
for the claim that such a feature
 is needed in order to obtain reasonably
expressive constraint solvers in a reasonably simple way
(see the discussion in \cite{Fruhwirth98}).
However this claim was not supported by any formal result, so far.

%On the other hand, the presence  of multiple heads complicates the semantics and the implementation of the language. In fact, obtaining a compositional semantics for CHR is much more complicated than for concurrent constraint programming and other similar languages (compare the models  in  \cite{BoePal91} and  in \cite{GM08}). Moreover,
%when considering the typical observables of CHR computations, no  fully  abstract model for CHR has been obtained, so far, while fully abstract semantics exist for several concurrent languages (see for example \cite{BoePal91,DGGab08,HoBaRu94,JR07}). Multiple heads complicate also the implementation since when a new constraint is placed in the ``store'', in order to find an applicable rule which can fire
%a multi-way join with selection has to be performed. In order to efficiently implement  this operation typically some static information is used by CHR optimizing compilers  \cite{HBJS,SD04}. The compiler for CHR described in \cite{SD04} can produce Prolog code which is as efficient as hand written Prolog programs, however in this case it is necessary to annotate the CHR source code by using type and mode declarations \cite{SSD05}. With this situation one could legitimately ask whether
%one can formally prove that multiple heads do indeed augment the expressive power of the language.

In this paper we prove that multiple heads do indeed augment the expressive power of CHR. Since we know that CHR is Turing powerful \cite{SnScDe05}, we first show  that CHR with single heads, called CHR$_1$ in what follows, is also Turing powerful, provided that the underlying constraint theory allows the equality predicate (interpreted as pattern matching) and that the signature contains at least one function symbol (of arity greater than zero). This result is certainly not surprising; however it is worth noting that, as we prove later, when considering an underlying (constraint theory defined over a) signature containing finitely many constant symbols and no function symbol CHR (with multiple heads) is still Turing complete, while this is not the case for CHR$_1$.

This provides a first separation result which is however rather weak, since usual constraint theories used in CHR do allow function symbols. Moreover, in general computability theory is not always the right framework for comparing the expressive power of  concurrent languages and several alternative formal tools have been proposed for this purpose.  In fact, most concurrent languages are Turing powerful and nevertheless, because of distributed and concurrent actions, they can exhibit a quite different observational behaviour and expressive power. For example, a language with synchronous communication allows to solve a distributed problem which is unsolvable by using the asynchronous version of that language \cite{Pal03}.

Hence, in the second part of the paper, we compare the expressive power of CHR and CHR$_1$ by using the notion of  language encoding, first formalized in \cite{BoePal94,Sh89,Vaan93}.
Intuitively, a language ${\cal L}$ is more expressive than a language ${\cal L}'$ or, equivalently, ${\cal L}'$  can be encoded in ${\cal L}$, if each program written in ${\cal L}'$ can be translated into an ${\cal L}$ program in such a way that: 1) the intended observable behaviour of the original program is preserved (under some suitable decoding); 2) the translation process satisfies some additional restrictions
%As discussed in \cite{BoePal94}, these additional properties are needed  in order to use the notion of encoding as a tool for language comparison. In fact, since the languages that we are considering are often Turing complete, they would always admit an encoding, provided that the observables for the target language are powerful enough.
%Therefore, as discussed in \cite{BoePal94},
%some further restrictions
%should be imposed on the decoder and on the compiler.
%Hence, in order to separate (Turing complete) languages in terms of encoding, often one adds some  further restrictions on the decoder (of the observables) and on the compiler (of programs).
which indicate how easy this process is and how
reasonable the decoding of the observables is. For example, typically one requires that the translation is compositional with respect to (some of) the syntactic operators of the language \cite{BoePal94}.
%that is, that ${\cal C}(A \ op\  B) =
%{\cal C}(A)\ op_s\ {\cal C}(B)$ holds, where  ${\cal C}$ is the compiler and $op_s$ is some counterpart of the operator $op$. This means that one can translate separately ``pieces'' of programs, which has important practical consequences.

%Since, differently from the case of process algebras, in CHR we have to distinguish a set of rules (the program) from the initial set of procedure calls (the goal), clearly we should adapt the above described technique for language comparison to our case.
%However such an adaptation is straightforward.

We prove that CHR cannot be encoded into CHR$_1$ under the following three assumptions. First we  assume that the observable properties to be preserved are the  constraints computed by a program for a goal, more precisely we consider data sufficient answers and qualified answers. Since these are the two typical CHR observables for most CHR reference semantics, assuming  their preservation is rather natural. Secondly we require that both the source CHR language and the target CHR$_1$ share the same  constraint theory defining built-in constraints.
This is also a natural assumption,
as CHR programs are usually written to define a new (user-defined) predicate in terms of the existing built-in constraints.
Finally we assume that the translation of a goal is compositional with respect to conjunction of goals, more precisely we assume  that $\encoding{A,B}_g$ = $\encoding{A}_g, \encoding{B}_g$ for any conjunctive goal $A,B$, where $\encoding{\ }_g$ denotes the translation of a goal. 
We believe this notion of compositionality to be reasonable as well, since  essentially it means that one can translate parts of the goal separately. It is worth noticing that we do not impose any restriction on the translation of the program rules.

%From this main separation result follows that CHR cannot be encoded in (constraint) logic programs nor in pure Prolog. This does not conflict with the fact that there exist many CHR to Prolog compilers: it simply means that these compilers do not satisfy our assumptions (typically, they do not translate goals in a compositional way).

%Our results are obtained firs considering data sufficient answers. Then  we consider the class of CHR programs which have qualified answers and trivial data sufficient answers only. Roughly, data sufficient answers are the results of terminating computations which contain built-in constraints only, while qualified answers can contain also some user-defined constraints, that is, some constraints which are defined by the program rules (rather than by the underlying theory). Trivial data sufficient answers are those identical to the original goal.
%We show that the previous separation result  holds also when considering this class of programs. This has some relevance because qualified answers are the main observable property  considered in the CHR semantics, and programs having qualified answers only are common.

Finally, our third contribution shows that also the number of atoms (greater than one) affects the expressive power of the language. In fact we prove that, under some slightly stronger assumptions on the translation of goals, there exists no encoding of CHR$_n$ (CHR with at most $n$ atoms in the head of the rules) into CHR$_m$, for $m<n$.

The remainder of this paper is organized as follows. Section \ref{chrsec:preliminaries} introduces the languages under consideration. We then provide the encoding of two counters machines \cite{Minsky67} in CHR$_1$ and discuss the Turing completeness of this language in Section \ref{chrsec:turing}.
Section \ref{chrsec:1vs2} contains the separation results for CHR and CHR$_1$ by considering first data sufficient answers and then qualified answers. In Section  \ref{chrsec:2vsn} we compare  CHR$_n$ and  CHR$_m$, while  Section \ref{chrsec:conclusions} concludes by discussing  related works and indicating some further development of this work.
%The proofs of the main results are contained in an Appendix which is included for the reviewer's convenience.
%
A shorter version of this paper, containing part of the results presented here, appeared in \cite{DGM08}.

\section{Preliminaries}\label{chrsec:preliminaries}
	
In this section we give an overview of CHR syntax and operational semantics following  \cite{Fruhwirth98}.
%Although the section is self-contained some familiarity with some basic notions of Constraint Logic Programming (\cite{JafMah94}) can be useful.

\subsection{CHR constraints and notation}

We first need to distinguish the constraints handled by an
existing solver, called built-in (or predefined) constraints, from
those defined by the CHR program, called user-defined (or CHR)
constraints. Therefore we assume that the signature $\Sigma$ on which programs are defined contains two disjoint sets of
predicate symbols $\Pi_b$ for built-in and $\Pi_u$ for user-defined constraints. In the following, as usual, an atomic constraint is a first-order atomic formula.

\begin{definition}[Built-in constraint]
 A \emph{built-in constraint} $c$ is defined
by: $$ c ::= a \ | \ c \wedge c \ | \ \exists_x c$$ where $a$ is an atomic constraint which uses a predicate symbol from $\Pi_b$.

%\footnote{We could
%consider more generally first order formulas as built-in
%constraints, as far as the results presented here are concerned.}.
 For built-in constraints we assume given a (first order) theory CT which
describes their meaning.
\end{definition}

\begin{definition}[User-defined constraint]
A \emph{user-defined (or CHR) constraint} is an ato\-mic constraint which uses a predicate symbols from $\Pi_u$.
%a multiset of atomic constraints which use predicate symbols from $\Pi_u$.
\end{definition}

%(note that in this case the multiplicity of atomic formulas is important, hence the conjunction has to be considered as a multiset).
We use $c,d$ to denote built-in
constraints, $h,k$ to denote CHR constraints and $a,b, f,g$ to denote
both built-in and user-defined constraints (we will call these
generally constraints).  The capital versions of these notations
will be used to denote multisets of constraints.
 We also denote by ${\tt false}$ any inconsistent conjunction of
constraints and with ${\tt true}$ the
empty multiset of built-in constraints.
%Furthermore we
%denote by ${\cal U}$ the set of user-defined constraints.
% and by
%${\cal B}$ the set of built-in constraints.

We will use ``,'' rather than $\wedge$ to denote conjunction
and we will often consider a conjunction of atomic constraints as
a multiset of atomic constraints: We prefer to use multisets rather than
sequences (as in the original CHR papers) because our results do not depend
on the order of atoms in the rules.
In particular, we will use this
notation based on multisets in the syntax of CHR.

The notation $\exists_{V} \phi$, where $V$ is a set of variables, denotes the
existential closure of a formula $\phi$ with respect to the variables in $V$, while the notation
$\exists_{-V} \phi$ denotes the
existential closure of a formula $\phi$ with the exception of the
variables in $V$ which remain unquantified. $Fv(\phi)$ denotes the
free variables appearing in $\phi$.
%and we denote by $\cdot$ the
%concatenation of sequences and by $\varepsilon$ the empty
%sequence. Given a set $A$,  $\wp(A)$ denotes the set consisting of all subsets of $A$, while
%$\wpm(A)$ denotes the set consisting of all the multisets over $A$.

Moreover, if $\bar t= t_1, \ldots t_m$ and $\bar t'= t'_1, \ldots t'_m$ are sequences of terms then the notation
$p(\bar t) = p'(\bar t')$
represents the set of equalities $\,t_1=t'_1, \ldots, t_m=t'_m\,$ if $p=p'$, and it is undefined otherwise. This notation is extended in the expected way to
multiset of constraints.
%Analogously, if $H=h_1, \ldots, h_k$ and $H'= h'_1, \ldots, h'_k$ are sequences of constraints, the notation
%$H=H'$ represents the set of equalities $h_1=h'_1, \ldots, h_k=h'_k$.
%Finally,  $\uplus$ denotes the multiset union, while
%we consider $\setminus$ as an overloaded operator used both for
%set and multiset difference (the meaning depends on the type of
%the arguments).
%Finally recall that a renaming is an invertible substitution which maps variables into variables, that is, %a substitution $\rho$ is a renaming if there exist a substitution $\rho^{-1}$ such that the composition %$\rho\rho^{-1}$ is the empty substitution.
%We follow the logic programming tradition and indicate the application of a substitution $\sigma$ to a %syntactic object $t$ by $t\sigma$.

\subsection{Syntax}
%We are now ready to introduce the CHR syntax as defined in \cite{Fruhwirth98}.
A CHR program is defined as a set of two kinds of rules: simplification and propagation (some papers consider also simpagation rules, since these are abbreviations for propagation
and simplification rules we do not need to introduce them). Intuitively, simplification rewrites constraints into simpler ones, while propagation adds new constraints which are logically redundant but may trigger further simplifications.

\begin{definition}\label{Syntax}
A CHR {\em simplification} rule has the form: $$ {\it r} \ @ \ H
\Leftrightarrow C \mid B$$
\noindent while a CHR {\em propagation} rule has the
form: $${\it r}\ @ \ H \Rightarrow C \mid B, $$
\noindent where ${\it r }$ is a
unique identifier of a rule, $H$ (the head) is a (non-empty) multiset of
user-defined constraints, $C$ (the guard) is a possibly empty multiset of built-in
constraints and $B$ (the body) is a possibly empty multiset of (built-in and
user-defined) constraints.

A CHR \emph{program} is a finite set of CHR simplification and propagation
rules.

A CHR {\em goal} is a multiset of (both user-defined and built-in) constraints.
\end{definition}

In the following, when the guard is $\texttt{true}$ we omit  $\texttt{true}  \,|\, $. Also the identifiers of rules are omitted when not needed.  An example of a CHR program is shown in next Section: Example \ref{ex:program}.

\subsection{Operational semantics}

We describe now the operational semantics of CHR by slightly modifying
the transition system defined in
\cite{Fruhwirth98}.

We use a transition system $T= ({\it Conf},\rrarrow)$ where configurations in ${\it
Conf}$ are triples of the form $\langle G,K,d \rangle$, where
$G$ are the constraints that remain to be solved, $K$ are the
user-defined constraints that have been accumulated and $d$ are
the built-in constraints that have been simplified. An {\em initial configuration} has the form
$\langle G,\emptyset,\emptyset \rangle$
%and consists of a goal $G$, an empty user-defined constraint and an empty built-in constraint.
while a  {\em final configuration} has either the form
$\langle G,K, \tt false \rangle$
when it is {\em failed},
%i.e. when it contains  an inconsistent
%built-in constraint store represented by the unsatisfiable
%constraint ${\tt false}$
or  the form
$\langle \emptyset,K,d \rangle$, where  $d$ is consistent, when it is successfully terminated because there are
no applicable rules.

%%\footnote{In
%\cite{Fruhwirth98} triples of the form $\langle G,K,d \rangle_{\cal V}$
%were used, where the annotation ${\cal V}$, which is not changed
%by the transition rules, is used to distinguish the variables
%appearing in the initial goal from the variables which are
%introduced by the rules. We can avoid such an indexing by
%explicitly referring to the original goal.}.

\begin{table}[tbp]
\tbl{The standard transition system for CHR \label{tor}}
{
\begin{tabular}{|ll|}  \hline
&
\\
\mbox{\bf Solve}& $\cfrac {\displaystyle   CT \models c \wedge d
\leftrightarrow d' \hbox{ and c is a built-in constraint}}
{\displaystyle \la(c,G), K,d\ra \rrarrow \la G, K,d'\ra
}$
\\
&
\\
\mbox{\bf Introduce}& $\cfrac {\displaystyle  \hbox{h is a
user-defined constraint}} {\displaystyle \la(h ,G), K,d\ra
\rrarrow \la G, (h,K),d\ra }$
\\
%&
%\\
%
&
\\
\mbox{\bf Simplify}& $\cfrac {\displaystyle   H \Leftrightarrow C
\mid B \in P \  \  x = Fv(H) \  \ CT \models d \rightarrow
\exists _{x} ((H=H')\wedge C)}  {\displaystyle \la G,H'\wedge
K,d\ra \rrarrow \la B\wedge G,K, H=H'\wedge d \ra }$
\\
&
\\
\mbox{\bf Propagate}& $\cfrac {\displaystyle   H \Rightarrow C \mid
B \in P \  \  x = Fv(H) \  \ CT \models d \rightarrow \exists
_{x} ((H=H')\wedge C) } {\displaystyle \la G,H'\wedge K,d\ra
\rrarrow \la B\wedge G,H'\wedge K, H=H'\wedge d \ra }$
\\
&
\\
\hline
\end{tabular}
}
\end{table}

Given a program $P$, the transition relation $\rrarrow
\subseteq {\it Conf} \times {\it Conf}$ is the least relation
satisfying the rules  in Table \ref{tor} (for the sake of
simplicity, we omit indexing the relation with the name of the
program).
The {\bf Solve} transition allows to update the
constraint store by taking into account a built-in constraint
contained in the goal.
%Without loss of generality, we will assume
%that $Fv(d') \subseteq Fv(c) \cup Fv(d)$.
The {\bf Introduce}
transition is used to move a user-defined constraint from the goal
to the CHR constraint store, where it can be handled by applying
CHR rules.

The transitions {\bf Simplify} and {\bf Propagate}
allow to rewrite user-defined constraints (which are in the CHR
constraint store) by using rules from the program. As usual, in
order to avoid variable name clashes, both these transitions
assume that
%clauses from the program are renamed apart, that is
%assumed that
all variables appearing in a program clause are fresh
ones. Both the {\bf Simplify} and {\bf Propagate} transitions are
applicable when the current store ($d$) is strong enough to entail
the guard of the rule ($C$), once the parameter passing has been
performed (this is expressed by the equation $H=H'$). Note that,
due to the existential  quantification over the variables $x$
appearing in $H$, in such a parameter passing the information flow
is from the actual parameters (in $H'$) to the formal parameters
(in $H$), that is, it is required that the constraints $H'$ which
have to be rewritten are an instance of the head $H$. This means that the equations $H=H'$ express pattern matching rather than unification.
%When
%applied, both these transitions add the body $B$ of the rule to
%the current goal and the set of equations $H=H'$ to the built-in constraint store.
The difference between {\bf Simplify} and {\bf Propagate} lies in the
fact that while the former transition removes the constraints $H'$
which have been rewritten from the CHR constraint store, this is
not the case for the latter.

Given a goal $G$, the operational semantics that we consider
observes the final store of computations terminating with an
empty goal and an empty user-defined constraint. 
We call these observables {\em data sufficient answers}
slightly deviating from the terminology of
\cite{Fruhwirth98} (a goal which has a data sufficient answer is called a data-sufficient
goal in \cite{Fruhwirth98}).

\begin{definition}\label{opsemsa}[Data sufficient answers]
Let $P$ be a program and let $G$ be a goal.  The set ${\cal
SA}_{P}(G)$ of data sufficient answers for the query $G$ in the
program $P$ is defined as: % follows
$${\cal SA}_P(G) = \{\exists_{-Fv(G)}  d \mid \la G, \emptyset,\emptyset \ra \rrarrow^{*} \la \emptyset, \emptyset, d \ra \nar \}.$$
\end{definition}

%Thus data sufficient answers consider the results of  terminated computations
%where all the user-defined constraints
%have been rewritten into built-in constraints. Following  \cite{Fruhwirth98}
We also consider the following different
notion of answer, obtained by computations terminating with a
user-defined constraint which does not need to be empty.

\begin{definition} \label{opsemqa}[Qualified answers \cite{Fruhwirth98}]
Let $P$ be a program and let $ G$ be a goal. The set ${\cal
QA}_{P}(G)$ of qualified answers for the query $G$ in the program
$P$ is defined as: % follows
$$ {\cal QA}_P(G) =
%\begin{array}[t]{l}
\{ \exists_{-Fv(G)} (K\wedge d )\mid \la G,
\emptyset,\emptyset \ra \rrarrow ^{*}\la \emptyset, K, d \ra
\nar \}.
% \\
% \cup
% \\
% \{ {\tt false} \mid \la G, \emptyset,\emptyset \ra
% \rrarrow ^*\la G', K, {\tt false} \ra   \}.
%\end{array}
$$
\end{definition}

Both previous
notions of observables characterize an input/output behaviour,
since the input constraint is implicitly considered in the goal.
Clearly in general ${\cal
SA}_{P}(G) \subseteq {\cal QA}_{P}(G)$ holds, since data sufficient answers
can be obtained by setting $K = \emptyset$ in Definition \ref{opsemqa}.

%In the remaining of this paper we will consider only
%simplification rules since propagation rules can be mimicked by
%simplification rules, as far as the results contained in this
%paper are concerned.

Note  that in the presence of propagation rules, the naive operational semantics that we consider here introduces
redundant infinite computations (because propagation rules do not
remove user defined constraints).
% (see rule \textbf{Propagate} in Table
%\ref{tor}), when a propagate rule is applied it introduces an
%infinite computation (obtained by subsequent applications of the
%same rule). Note however that this does not imply that in presence
%of an active propagation rule the semantics that we consider is
%empty. In fact, the application of a simplification rule after a
%propagation rule can cause the termination of the computation, by
%removing the atoms which are needed by the head of the propagation
%rule.
It is  possible to define a different operational
semantics (see \cite{Ab97}) which avoids these
infinite computations by allowing to apply at most once a
propagation rule to the same constraints. The results presented here hold also in case this semantics is considered, essentially because the number of applications of propagations rules does not matter. We refer here to the naive operational semantics because it is simpler than that one in \cite{Ab97}.

An example can be useful to see what kind of programs  we are considering here. The following program implements the sieve of Eratosthenes to compute primes.% and it is one of the very first examples of CHR usage.

\begin{example}\label{ex:program}
The following CHR program which consists of three simplifications rules, given a goal $upto(N)$ with $N$ natural number, computes all prime numbers up to $N$: the first two rules generate all the possible candidates as prime numbers, while the third one removes all the incorrect information.
\begin{align*}
& upto(1) \Leftrightarrow \true\\
& upto(N) \Leftrightarrow N>1 \mid prime(N), upto(N-1)\\
& prime(X), prime(Y) \Leftrightarrow X \ {\tt mod} \ Y = 0 \mid prime(Y).
\end{align*}
For example suppose that the goal is $upto(4)$ then the following is one of the possible computations of the program where, by using the first two rules, from the goal $upto(4)$ we can generate all possible candidates,
$$\tuple{upto(4), \emptyset, \emptyset} \rrarrow^{*} \tuple{\emptyset, (prime(4), prime(3), prime(2)), \emptyset}.$$
Then the third rule can be used to check, for every couple of constraints $prime(X)$, $prime(Y)$, if $X$ is divisible by $Y$ and in this case restores in the pool of constraints only the constraint $prime(Y)$ (in other words, we remove the constraint $prime(X)$). Thus we obtain:
$$\tuple{\emptyset, (prime(4), prime(3), prime(2)), \emptyset} \rrarrow^{*} \tuple{\emptyset, ( prime(3), prime(2)), \emptyset}.$$
Since there are no applicable rules $\tuple{\emptyset, ( prime(3), prime(2)), \emptyset}$ is a final configuration. Note that this is a qualified answer and the program with this goal has no data sufficient answers.
\end{example}

In the following, we study several CHR dialects defined by setting a limit in the number of the atoms present in the head of rules and by considering the possibility of non trivial data sufficient answers, as described by the following two definitions.

\begin{definition}\label{def:tnt}
A data sufficient answer for the goal $G$ in the program $P$ is called  trivial  if it is equal to $G$ (is called non trivial otherwise).
\end{definition}

\begin{definition}[CHR dialects]\label{def:dialects}
With CHR$_n$ we denote a CHR language where the number of atoms in the head of the rules is at most $n$.
Moreover, CHR$_{n,d}$ denotes the language consisting of CHR$_n$  programs which have (for some goal) non trivial data sufficient answers, while CHR$_{n,t}$ denotes the language consisting of CHR$_n$ programs which, for any goal, have only trivial data sufficient answers and qualified answers.
\end{definition}

\section{On the Turing completeness of CHR$_1$}\label{chrsec:turing}
 	In this section we discuss the Turing completeness of CHR$_1$ by taking into account also the underlying constraint theory CT and signature $\Sigma$ (defined in the previous section). In order to show the Turing completeness of a language we encode two counter machines, also called Minsky machines, into it.

We recall here some basic notions on this Turing equivalent formalism. A two counter machine (2CM) \cite{Minsky67} $M(v_0, v_1)$ consists of two registers $R_1$ and $R_2$ holding arbitrary large natural numbers and initialized with the values $v_0$ and $v_1$, and a program, i.e. a finite sequence of numbered instructions which modify the two registers.
There are  three types of instructions $j:Inst()$ where $j$ is the number of the instruction:
\begin{itemize}
 \item $j:Succ(R_i)$: adds 1 to the content of register $R_i$ and goes to instruction $j+1$;
 \item $j:DecJump(R_i, l)$: if the content of the register $R_i$ is not zero, then decreases it by 1 and goes to instruction $j+1$, otherwise jumps to instruction $l$;
 \item $j:Halt$: stops computation and returns the value in register $R_1$,
\end{itemize}
where $1\leq i \leq 2$; $1 \leq j,l \leq n$ and $n$ is the number of instructions of the program.

An internal state of the machine is given by a tuple $(p_i, r_1, r_2)$ where the program counter $p_i$ indicates the next instruction and $r_1$, $r_2$ are the current contents of the two registers. Given a program, its computation proceeds by executing the instructions as indicated by the program counter. The execution stops when the program counter reaches the $Halt$ instruction.

As a first result, we show that CHR$_1$ is Turing powerful, provided that the underlying language signature $\Sigma$ contains at least a function symbol (of arity one) and a  constant symbol. This result is obtained by  providing an encoding $\encoding{\ }: Machines \rightarrow {\cal P}_{1}$ of a  two counter machine  $M(v_0, v_1)$ in a CHR program (${\cal P}_{1}$ denotes the set of CHR$_1$ programs) as shown in Figure \ref{tab:Minsky}: Every rule takes as input the program counter and the two registers and updates the state according to  the instruction in the obvious way.

Note that, due to the pattern matching mechanism, a generic goal
$i(p_i, s, t)$ can fire at most one of the two rules encoding the $DecJump$ instruction (in fact, if $s$ is a free variable no rule in the encoding of
 $p_i: DecJump(r_1, p_l)$ is fired).
\begin{figure}[t]
 \begin{center}
\fbox{
\begin{minipage}{12.2cm}
 \begin{align*}
 &\encoding{p_i:Succ(r_1)} := && i(p_i, R_1, R_2) \Leftrightarrow i(p_{i+1}, succ(R_1), R_2)\\
 &\encoding{p_i:Succ(r_2)} := && i(p_i, R_1, R_2) \Leftrightarrow i(p_{i+1}, R_1, succ(R_2))\\
 &\encoding{p_i:DecJump(r_1, p_l)} := && i(p_i, 0, R_2) \Leftrightarrow  i(p_l, 0, R_2)\\														 &&& i(p_i, succ(R_1), R_2) \Leftrightarrow    i(p_{i+1}, R_1, R_2)\\
 &\encoding{p_i:DecJump(r_2, p_l)} := && i(p_i, R_1, 0) \Leftrightarrow   i(p_l, R_1, 0)\\
	&&& i(p_i, R_1, succ(R_2)) \Leftrightarrow    i(p_{i+1}, R_1, R_2)
\end{align*}
\end{minipage}
}
 \end{center}
\caption{2CM encoding in CHR$_1$}\label{tab:Minsky}
\end{figure}

Without loss of generality we can assume that the counters are initialized with $0$, hence the encoding of a machine $M$ with $n$ instructions has the form:
$$ \encoding{M(0,0)} := \{ \encoding{Instruction_1}, \dots,  \encoding{Instruction_n}\}$$
(note that the initial values of the registers are not considered in the encoding of the machine: they will be used in the initial goal, as shown below). The following theorem, whose proof is immediate, states the correctness of the encoding (we use the notation
$succ^k(0)$ to denote $k$ applications of the functor $succ$ to $0$).

\begin{theorem}\label{th1}
 A 2CM $M(0,0)$ halts with output $k>0$ (or $k=0$) on register $R_1$ if and only if the goal $i(1, 0, 0)$ in the program $\encoding{M(0,0)}$ has a qualified answer $i(p_j, R'_1, R'_2)$, where $R'_1=succ^k(0)$ (or $R'_1=0$).
\end{theorem}

Note that the encoding provided in Figure \ref{tab:Minsky} does not use any built-in, hence we can consider an empty theory CT here\footnote{We used the = built-in the the operational semantics in order to perform parameter passing, however this is only a meta-notation which does not mean that the built-in = has to be used in the language.}.
If the = built-in is allowed in the body of rules then one could provide an encoding which gives the results of computation in terms of data sufficient answer, rather than qualified answer. To obtain this it is sufficient to add a fourth argument  $X$ (for the result) to the predicate $i$ and to add the following translation for the $Halt$ instruction:
$$
 \encoding{p_i:Halt} :=  i(p_i, R_1, R_2, X) \Leftrightarrow X=R_1.
$$
 Such a translation in the previous encoding was not needed, since when one find the $Halt$ instruction the CHR program simply stops and produces a qualified answer.

It is also worth noting that the presence of a function symbol ($succ()$ in our case)  is crucial in order to encode natural numbers and therefore to obtain the above result.
Indeed, when considering a signature containing only a finite number of constant symbols the language CHR$_1$, differently from the case of CHR, is not Turing powerful. To be more precise, assume that CT defines only the = symbol, interpreted as pattern matching,  which cannot be used in the body of rules  (it can be used in the guards only). Assume also that the CHR language is now defined over a signature $\Sigma$ containing finitely many constant symbols and no function symbol (of arity $>$ 0).
Let us call  $CHR_\emptyset$ the resulting language.

As observed in \cite{Sn08}, $CHR_\emptyset$ (called in that paper single-headed CHR without host language) is computationally equivalent to propositional CHR (i.e. CHR with only zero-arity constraints), which can easily encoded into Petri nets. Since it is well known that in this formalism termination is decidable, we have the following result.

\begin{theorem}\label{th:turing}  \cite{Sn08}
 CHR$_\emptyset$  is not Turing complete.
\end{theorem}

On the other hand, CHR (with multiple heads) is still Turing powerful also when considering a signature containing finitely many constant symbols and no function symbol, and assuming that CT defines only the = symbol  which is interpreted as before and which, as before, cannot be used in the body of rules.  Indeed, as we show in Figure \ref{tab:Minsky2}, under these hypothesis we can encode 2CMs into CHR. The basic idea of this encoding is that to represent the values of the registers we use chains (conjunctions) of atomic formulas of the form
$s(R_1, SuccR_1)$, $s(SuccR_1, SuccR_1') \ldots$ (recall that $R_1$, $SuccR_1$,
$SuccR_1'$ are variables and we have countably many variables; moreover recall that the CHR computation mechanism avoids variables capture by using fresh names for variables each time a rule is used).

As we discuss in the conclusions this encoding, suggested by Jon Sneyers in a review of a previous version of this paper, is similar to those existing in the field of concurrency theory. Nevertheless, there are important technical differences. In particular, it is worth noting that for the correctness of the encoding it is essential that pattern matching rather than unification is used when applying rules: In fact this ensures that in the case of the decrement only one of the two instructions can match the goal and therefore can be used. The correctness of the encoding is stated by the following theorem whose proof is immediate.

\begin{figure}[t]
 \begin{center}
\fbox{
\begin{minipage}{12.2cm}
 \begin{align*}
 &\encoding{p_i:Succ(r_1)}_2 := && i(p_i, R_1, R_2) \Leftrightarrow s(R_1,SuccR_1), i(p_{i+1}, SuccR_1, R_2)\\
 &\encoding{p_i:Succ(r_2)}_2 :=
&&i(p_i, R_1, R_2) \Leftrightarrow s(R_2,SuccR_2), i(p_{i+1}, R_1,  SuccR_2)\\
 &\encoding{p_i:DecJump(r_1, p_l)}_2 :=
&& i(p_i, R_1, R_2), s(PreR_1, R_1) \Leftrightarrow  i(p_{i+1}, PreR_1, R_2)\\
& &&  zero(R_1), i(p_i, R_1, R_2) \Leftrightarrow i(p_l, R_1, R_2), zero(R_1)\\
 &\encoding{p_i:DecJump(r_2, p_l)}_2 := &&  i(p_i, R_1, R_2), s(PreR_2, R_2) \Leftrightarrow  i(p_{i+1}, R_1, PreR_2)\\
& & &  zero(R_2), i(p_i, R_1, R_2) \Leftrightarrow i(p_l, R_1, R_2), zero(R_2)
\end{align*}
\end{minipage}
}
 \end{center}
\caption{2CM encoding in CHR}\label{tab:Minsky2}
\end{figure}

\begin{theorem}\label{th:turing2}
A 2CM $M(0,0)$ halts with output $k>0$ (or $k=0$) if and only if the goal $zero(R_1)\wedge zero(R_2)\wedge i(1, R_1, R_2)$ in the program $\encoding{M(0,0)}_2$ produces a qualified answer
$$\begin{array}{ll}
    \vspace*{0.1cm}\exists_{- R_1,R_2}(\begin{array}[t]{l}
    i(p_j, SuccR_1^{k}, R'_2) \wedge  zero(R_1) \wedge s(R_1, SuccR_1^1) \wedge \\
    \vspace*{0.1cm}\bigwedge_{i= 1}^{k-1} s(SuccR_1^{i}, SuccR_1^{i+1})\wedge H),
     \end{array} \\
    \mbox{where } Fv(H) \cap \{R_1, SuccR_1^1, \ldots, SuccR_1^{k} \} = \emptyset
  \end{array}
$$
 (or $\exists_{- R_1, R_2}(i(p_j, R_1, R'_2) \wedge zero(R_1) \wedge H)$, where
$Fv(H) \cap \{R_1\} = \emptyset$).\end{theorem}

Previous Theorems state a separation result between CHR and CHR$_1$, however this is rather weak since the real implementations of CHR usually consider a non-trivial  constraint theory which includes function symbols. Therefore we are interested in proving finer separation results which hold for Turing powerful languages.  This is the content of the following section.

\section{Separating CHR and CHR$_1$}\label{chrsec:1vs2}

%\todo{file: expressiveness.tex}

In this section we consider a generic constraint theory CT  which allows the built-in predicate $=$ and we assume that the signature contains at least a constant and a function (of arity $ >$ 0) symbol. We have seen that in this case both CHR and CHR$_1$ are Turing powerful, which means that in principle one can always encode CHR into CHR$_1$. The question is how difficult and how acceptable such an encoding is and this question can have important practical consequences: for example, a distributed algorithm can be implemented in one language in a reasonably simple way and cannot be implemented in another (Turing powerful) language, unless one introduces rather complicated data structures or loses some compositionality properties (see \cite{VPP07}).

We prove now that, when considering \emph{acceptable encodings} and generic goals (whose components can share variables) CHR cannot be embedded into CHR$_1$  while preserving data sufficient answers. As a corollary we obtain that also qualified answers cannot be preserved. This general result is obtained by proving two more specific results.

First we have to formally define what an acceptable encoding is. We do this by giving a generic definition, which will be used also in the next section, which considers separately program and goal encodings. Hence in the following we denote by CHR$_x$ some CHR (sub)language and assume that ${\cal P}_{x}$ is the set of all the CHR$_x$ programs while ${\cal G}_{x}$ is the set of possible CHR$_x$ goals. Usually the sub-language is defined by suitable syntactic restrictions, as in the case of CHR$_1$, however in some cases we will use also a semantic characterization, that is, by a slight abuse of notation, we will identify a sub-language with a set of programs having some specific semantic property.
A  \emph{program encoding} of CHR$_x$ into CHR$_y$ is then defined as any  function $\encoding{\ }:{\cal P}_{x} \rightarrow {\cal P}_{y}$. To simplify the treatment we assume that both the source and the target language of the program encoding use the same built-in constraints semantically described by a theory CT. Note that we do not impose any other restriction on the program translation (which, in particular, could also be non compositional).

Next we have to define how the initial goal of the source language has to be translated into the target language. Analogously to the case of programs, the goal encoding is a function $\encoding{\ }_g:{\cal G}_{x} \rightarrow {\cal G}_{y}$, however here we require that such a function is compositional (actually, an homomorphism) with respect to the conjunction of atoms, as mentioned in the introduction. Moreover, since both the source and the target language share the same constraint theory, we assume that the built-ins present in the goal are left unchanged. These assumptions  essentially mean that our encoding respects the structure of the original goal and does not introduce new relations among the variables which appear in the goal.
Note that we differentiate the goals ${\cal G}_{x}$ of the source language from those  ${\cal G}_{y}$ of the target one because, in principle, a CHR$_y$ program could use some user defined predicates which are not allowed in the goals of the original program -- this means that the signatures of (language of) the original and the translated program could be different.
Note also that the following definitions are parametric  with respect to a class ${\cal G}$ of goals: clearly considering different classes of goals could affect our encodability results. Such a parameter will be instantiated when the notion of acceptable encoding will be used.

Finally, as mentioned before, we are interested in preserving data sufficient or qualified answers.  Hence we have the following definition.

\begin{definition}[Acceptable encoding ]\label{def:resenc}
Let ${\cal G}$ be  a class of CHR goals and let CHR$_x$ and CHR$_y$ be two CHR (sub)languages. An {\em acceptable encoding} of CHR$_x$ into CHR$_y$, for the class of goals ${\cal G}$, is a pair of mappings $\encoding{\ } :{\cal P}_{x} \rightarrow {\cal P}_{y}$ and $\encoding{\ }_g:{\cal G}_{x} \rightarrow {\cal G}_{y}$ which satisfy the following conditions:
\begin{enumerate}
 \item ${\cal P}_{x}$ and ${\cal P}_{y}$  share the same CT;
 \item for any goal $(A,B) \in {\cal G}_{x}$, $\encoding{A,B}_g$ = $\encoding{A}_g, \encoding{B}_g$ holds. We also assume that the built-ins present in the goal are left unchanged;
 \item Data sufficient answers are preserved for the set of programs ${\cal P}_{x}$ and the class of goals ${\cal G}$, that is, for all $P\in {\cal P}_{x}$ and $G\in {\cal G}$,
 ${\cal SA}_{P}(G) =  {\cal SA}_{\encoding{P}}(\encoding{G}_g)$.
%\footnote{We will consider separately the two cases of data sufficient and qualified answers.}
\end{enumerate}
Moreover we define an  {\em acceptable encoding for qualified answers} of CHR$_x$ into CHR$_y$, for the class of goals ${\cal G}$, exactly as an acceptable encoding, with the exception that the third condition above is replaced by the following:
\begin{description}
 \item[(3')] Qualified answers are preserved for the set of programs ${\cal P}_{x}$ and the class of goals ${\cal G}$, that is, for all $P\in {\cal P}_{x}$ and  $G\in {\cal G}$,
 ${\cal QA}_{P}(G) =  {\cal QA}_{\encoding{P}}(\encoding{G}_g)$.
%Moreover an encoding is said to be {\em weakly acceptable} when the last condition is substituted for the following:
%\begin{itemize}
%\item Data sufficient (qualified) answers are weakly preserved for the class of goals ${\cal G}$, that is, for all $G\in {\cal G}$, ${\cal SA}_{P}(G) \subseteq  {\cal SA}_{\encoding{P}}(\encoding{G}_g)$
% (${\cal QA}_{P}(G) \subseteq {\cal QA}_{\encoding{P}}(\encoding{G}_g)$) holds.
% \end{itemize}
\end{description}
\end{definition}

%
%\begin{definition}[Acceptable encoding w.r.t. Qualified Answers]\label{def:resencqa}
%Let ${\cal G}$ be  a class of CHR goals. An {\em acceptable encoding} (of CHR into CHR$_1$, for the class of goals ${\cal G}$) is a pair of mappings $\encoding{\ } :{\cal P}_{CHR} \rightarrow {\cal P}_{CHR_1}$ and $\encoding{\ }_g:{\cal G}_{CHR} \rightarrow {\cal G}_{CHR_1}$ which satisfy the following conditions:
%\begin{itemize}
% \item ${\cal P}_{CHR}$ and ${\cal P}_{CHR_1}$  share the same CT;
% \item for any goal $(A,B) \in {\cal G}_{CHR}$, $\encoding{A,B}_g$ = $\encoding{A}_g, \encoding{B}_g$ holds. We also assume that the built-ins present in the goal are left unchanged;
% \item Qualified answers are preserved for the class of goals ${\cal G}$, that is, for all $G\in {\cal G}$, %\subseteq {\cal G}_{CHR}$,
% ${\cal QA}_{P}(G) =  {\cal QA}_{\encoding{P}}(\encoding{G}_g)$ holds.
%\end{itemize}
%\end{definition}

%

%

Obviously the notion of acceptable encoding for qualified answers is stronger than that one of acceptable encoding, since  ${\cal SA}_{P}(G) \subseteq  {\cal QA}_{P}(G) $ holds.
Note also that, since we consider goals as multisets, with the second condition in the above definition we are not requiring that the order of atoms in the goals is preserved by the translation: We are only requiring that
the translation of $A, B$ is the conjunction of the translation of $A$ and of $B$, i.e. the encoding is homomorphic.
Weakening this condition by requiring that the translation of $A, B$ is some form of composition of the translation of $A$ and of $B$ does not seem reasonable, as conjunction is the only form for goal composition available in these languages.
Moreover, homomorphic encoding are a quite common assumption in the papers studying expressiveness of concurrent languages, see for example \cite{Pal03}.

We are now ready to prove our separation results. Next section considers only data sufficient answers.

%We need now to define more precisely what it means for a goals to be share-free
%(and to share variables).

%\begin{definition}\label{def:fresharegoals}
% Consider a goal $ (c, H,G)$ where c is a built-in constraint and $H$ and $G$
%are (multisets of) user defined predicates.
% Such a goal is called share-free iff \begin{itemize}
% \item $H$ and $G$ do not share variables;
%  \item $c= c_1 \wedge c_2$ and $CT \models c \leftrightarrow \exists_
%{-Fv(H)}c_1 \wedge \exists_{-Fv(G)}c_2$.
%  \end{itemize}
%  If a goal is not share-free it is called sharing goal.
%  \end{definition}

%Clearly the most interesting goals are those which are sharing ones, since (direct or indirect) sharing of variables is a way to express parameter passing among processes.

\subsection{Separating CHR and CHR$_1$ by considering data sufficient answers}

In order to prove our first separation result we need the following lemma which states two key properties of CHR$_1$ computations.
The first one says that if the conjunctive $G,H$ with input constraint $c$ produces a data sufficient answer $d$, then when considering one component, say $G$, with the input constraint $d$
we obtain the same data sufficient answer.  The second one states that when considering the subgoals $G$ and $H$ there exists at least one of them which allows to obtain the same data sufficient answer $d$ also starting with an input constraint $c'$ weaker than $d$.

%
%\revcinzia{Correzioni come da rev. 2 controllare anche either or}
%\begin{lemma*}\label{lem:preservarisposte}
%\textcolor{red}{VECCHIO}
% Let $P$ be a CHR$_1$ program  and let $ (c, G, H)$ be a goal, where $c$ is a built-in constraint, $G$ and $H$ are multisets of CHR constraints and $V=Fv(c,G,H)$. Assume that $ (c, G, H)$ in $P$ has the data sufficient answer $d$.
%Then the following holds:
%\begin{itemize}
%\item
%Both the goals $(d, G)$  and
%$(d, H)$ have the same data sufficient answer $d$.
%\item If $\ CT \models c \not \rightarrow d$ then there exists a built-in constraint $c'$ such that $Fv(c') \subseteq V$, $CT \models c' \not \rightarrow d$ and
%either $(c', G)$ or $(c', H)$ has the data sufficient answer $d$.
%\end{itemize}

%\end{lemma*}

\begin{lemma}\label{lem:preservarisposte}
 Let $P$ be a CHR$_1$ program  and let $ (c, G, H)$ be a goal, where $c$ is a built-in constraint, $G$ and $H$ are multisets of CHR constraints. Let $V=Fv(c,G,H)$ and assume that $ (c, G, H)$ in $P$ has the data sufficient answer $d$.
Then the following holds:
\begin{itemize}
\item
Both the goals $(d, G)$  and
$(d, H)$ have the same data sufficient answer $d$.
\item If $\ CT \models c \not \rightarrow d$ then there exists a built-in constraint $c'$ such that $Fv(c') \subseteq V$, $CT \models c' \not \rightarrow d$ and
at least one of the two goals  $(c', G)$ and $(c', H)$ has the data sufficient answer $d$.
\end{itemize}

\end{lemma}

\begin{proof}
The proof of the first statement is straightforward (since we consider single headed programs). In fact, since the goal $(c,G,H)$ has the data sufficient answer $d $ in $P$, the goal $(d,G)$ can either answer $d$ or can produce a configuration where the user defined constraints are waiting for some guards to be satisfied in order to apply a rule $r$, but since the goal contains all the built-in constraints in the answer all the guards are satisfied letting the program to answer $d$.

We prove the second statement.
Let  $$ \delta=\la (c, G, H), \emptyset,\emptyset \ra
\rrarrow ^{*}\la \emptyset, \emptyset, d' \ra \nar$$ be the derivation producing the data sufficient answer $d=\exists _{-V} d'$ for the goal $(c, G, H)$.

By definition of derivation and since by hypothesis $\ CT \models c \not \rightarrow d$,
$\delta$ must be of the form
 $$  \la (c, G, H), \emptyset,\emptyset \ra \rrarrow ^{*}\la (c_1, G_1) , S_1, d_1 \ra
      \rrarrow
            \la (c_2, G_2) , S_2, d_2 \ra
      \rrarrow ^{*}
      \la \emptyset, \emptyset, d' \ra \nar,
     $$
where for $i\in [1,2]$, $c_i$ and $d_i$  are built-in constraints such that
$CT \models c_1 \wedge d_1 \not \rightarrow d$ and
$CT \models c_2 \wedge d_2 \rightarrow d$.
We choose $c'= \exists _{-V} (c_1 \wedge d_1)$.
By definition of derivation and since $P$ is a CHR$_1$ program, the transition
$$\la (c_1, G_1) , S_1, d_1 \ra \rrarrow
\la (c_2, G_2) , S_2, d_2 \ra$$ must be a rule application of a single headed rule $r$, which must match with a constraint $k$ that was derived (in the obvious sense) by $G$ or $H$. Without loss of generality, we can assume that $k$ was derived from $G$.
By construction $c'$ suffices to satisfy the guards needed to reproduce $k$, which can then fire the rule $r$, after which all the rules needed to let the constraints of $G$ disappear can fire.
Therefore we have that $$\la (c', G) , \emptyset, \emptyset \ra \rrarrow ^{*}
 \la \emptyset, \emptyset, d'' \ra \nar$$ where $CT \models \exists _{-V} d''\leftrightarrow \exists _{-V} d' (\leftrightarrow d)$ and then the thesis follows.
\end{proof}

 Note that Lemma \ref{lem:preservarisposte} is not true anymore if we consider (multiple headed) CHR programs.
 Indeed if we consider the program $P$ consisting of the single rule $$\texttt{rule}\ @\ H, H \Leftrightarrow \true \mid c$$ then the goal $(H, H)$ has the data sufficient answer $c$ in $P$, but for each constraint $c'$ the goal  $(H, c')$ has no data sufficient answer in $P$.
With the help of the previous lemma we can now prove our main separation result.
The idea of the proof is that any possible encoding of the  rule $$ \simp{r}{H, G}{\true}{c}$$ into CHR$_1$ would either produce more answers for the goal $H$ (or $G$), or would not be able to provide the answer $c$ for
the goal $H,G$.
Using the notation introduced in Definition \ref{def:dialects} and considering $\subseteq$ as multiset inclusion, we have then the following.

\begin{theorem}\label{th:general}
Let ${\cal G}$ be a class of goals such that if  $H$  is a head of a rule then $K \in {\cal G}$ for any $K\subseteq H$. Then, for n$\geq 2$, there exists no acceptable encoding of
CHR$_{n,d}$ in CHR$_1$ for the class ${\cal G}$.
\end{theorem}
\begin{proof}
%According to Definition \ref{def:resenc} to prove the theorem it is enough to exhibit a CHR program which cannot be embedded in CHR$_1$, while preserving data sufficient answers for a goal.
 The proof is by contradiction. Assume that there exists an acceptable encoding
$\encoding{\ } :{\cal P}_{n,d} \rightarrow {\cal P}_{1}$ and $\encoding{\ }_g:{\cal G}_{n,d} \rightarrow {\cal G}_{1}$
 of CHR$_{n,d}$ into CHR$_1$ for the class of goals ${\cal G}$ and let $P$ be the program consisting of the single rule
 $$ \simp{r}{H, G}{\true}{c}.$$
%Note that $H$ and $G$ here can share variables, hence we can consider
%the class of sharing goals.
Assume also, that $c$ (restricted to the variables in $H,G$) is not the weakest constraint, i.e. assume that there exists  $d$  such that $CT \models d \not \rightarrow \exists _{-V} c$ where $V = Fv(H,G)$. Note that this assumption does not imply any loss of generality, since, as mentioned at the beginning of this section, we assume that the constraint theory allows the built-in predicate $=$ and the signature contains at least a constant and a function (of arity $>$ 0) symbol.

Since  the goal   $(H, G)$ has the data sufficient answer  $\exists _{-V} c$ in the program $P$ and since the encoding preserves data sufficient answers,  the goal $\encoding{ (H, G)}_g$ has the data sufficient answer  $\exists _{-V} c$ also in the program $\encoding{P}$.  From the compositionality of the translation of goals and the previous Lemma \ref{lem:preservarisposte} it follows that there exists a constraint $c'$ such that $Fv(c') \subseteq V$,
$CT \models c'\not \rightarrow \exists _{-V} c$ and
at least one of the two goals $\encoding{(c',H)}_g,$
 and $\encoding{ (c', G) }_g$ has the data sufficient answer $c$ in the encoded program $\encoding{P}$.

However neither  $(c', H)$
 nor $(c', G)$ has any  data sufficient answer in the original program $P$. This contradicts the fact that
 there exists an acceptable encoding
 of CHR$_{n,d}$ into CHR$_1$ for the class of goals ${\cal G}$, thus concluding the proof.
%The thesis for qualified answers follows immediately from the previous part, as  qualified answers contain the set of data sufficient answers.
\end{proof}

Obviously, previous theorem implies that (under the same hypothesis) no acceptable encod\-ing for qualified answers of CHR$_{n,d}$ into CHR$_1$ exists,
since  ${\cal SA}_{P}(G) \subseteq  {\cal QA}_{P}(G)$. The hypothesis we made on the class of goals ${\cal G}$ is rather weak, as typically heads of rules have to be used as goals.
From  Theorem~\ref{th:general} we have immediately the
following.

\begin{corollary}\label{ch:general}
Let ${\cal G}$ be a class of goals such that if  $H$  is a head of a rule then $K \in {\cal G}$ for any $K\subseteq H$. Then, for n$\geq 2$, there exists no acceptable encoding (for qualified answers) of
CHR$_{n}$ in CHR$_1$ for the class ${\cal G}$.
\end{corollary}

As an example of the application of the previous theorem consider the program (from  \cite{Fruhwirth98}) contained
%In \cite{Fruhwirth98} the author claims that he can build a program where multiple heads are necessary, since there is no proof of this result, here we want to formally justify his claim.
in Figure \ref{tab:lessequal} which allows one to define the user-defined constraint \textit{Lessequal} (to be interpreted as  $\leq$) in terms of the only  built-in constraint = (to be interpreted as syntactic equality).
\begin{figure}[t]
\begin{center}
\fbox{
\begin{minipage}{12cm}
\begin{align*}
&\simp{reflexivity}{Lessequal(X,Y)}{X=Y}{\true}\\
&\simpsenza{antisymmetry}{Lessequal(X,Y), Lessequal(Y,X)}{X=Y}\\
&\propsenza{transitivity}{Lessequal(X,Y), Lessequal(Y,Z)}{}{Lessequal(X,Z)}
\end{align*}
      \end{minipage}
}
\end{center}
\caption{A program for defining  $\leq$  in CHR}\label{tab:lessequal}
\end{figure}
For instance, given the goal $\{ Lessequal(A,B), Lessequal(B,C),$ $Lessequal(C,A)\}$
 after a few computational steps the program will answer $A=B, B=C, C=A$.
 Now, for obtaining this behaviour, it is essential to use multiple heads, as already claimed in \cite{Fruhwirth98} and formally proved by the previous theorem.
In fact, following the lines of the proof of Theorem  \ref{th:general}, one can show that
if a single headed program $P'$ is  any translation of the program in Figure \ref{tab:lessequal} which produces the correct answer for the goal above, then there exists a subgoal which has an answer in $P'$ but not in the original program.

\subsection{Separating CHR and CHR$_1$ by considering qualified answers}

Theorem \ref{th:general} assumes that programs have non trivial data sufficient answers. Nevertheless, since qualified answers are the most interesting ones for CHR programs, one could wonder what happens when considering the CHR$_{n,t}$ language (see Definition \ref{def:dialects}).

Here we prove that also CHR$_{n,t}$  cannot be encoded into CHR$_1$.  The proof of this result is somehow easier to obtain since the multiplicity of atomic formulae here is important. In fact,  if $u(x,y)$ is a user-defined constraint,
the meaning of $u(x,y)$, $u(x,y)$ does not necessarily coincide with that one of $u(x,y)$. This is well known also in the case of logic programs (see any article on the S-semantics of logic programs): consider, for example, the  program:
$$  u(x,y) \Leftrightarrow  x = a \quad  \quad  \quad
%\\
u(x,y) \Leftrightarrow  y = b$$
%\end{array}
%\]
which is essentially a pure logic program written with the CHR syntax. Notice that when considering an abstract operational semantics, as the one that we consider here, the presence of commit-choice does not affect the possible results. For example, in the previous program when reducing the goal $u(x,y)$ one can always choose (non deterministically) either the first or the second rule.
% Of course in CHR  there is the commit-choice which is not present  in pure logic programming. However notice that when considering an abstract operational semantics, as the one that we consider here, the presence of commit-choice does not affect the possible results. For example, in the previous program when reducing the goal $u(x,y)$ one can always choose (non deterministically) either the first or the second rule.}.

Now the goal $u(x,y),u(x,y)$ in such a program has the (data sufficient) answer $x=a, y = b$  while this is not the case for the goal $u(x,y)$ which has the answer  $x=a$ and the answer $y = b$ (of course, using guards one can make more significant examples). Thus, when considering user-defined predicates, it is acceptable to distinguish $u(x,y),u(x,y)$  from $u(x,y)$, i.e. to take into account the multiplicity. This is not the case for  ``pure'' built-in constraints, since the meaning of a (pure) built-in is defined by a first order theory CT in terms of logical consequences,  and from this point of view $b\wedge b$ is equivalent to $b$.

In order to prove our result we need first the following result which states that, when considering single headed rules, if the goal is replicated then there exists a computation where at every step a rule is applied twice. Hence it is easy to observe that if the computation will terminate producing a qualified answer which contains an atomic user-defined constraint, then such a constraint is replicated. More precisely we have the following Lemma whose proof is immediate.

\begin{lemma}\label{lem:single}
 Let $P$ be a CHR$_1$ program. If $(G, G)$ is a goal whose evaluation in $P$ produces a qualified answer $(c, H)$ containing the atomic user-defined constraint $k$, then the goal $(c, G, G)$ has a qualified answer containing $(k,k)$.
\end{lemma}

Hence we can prove the following separation result.
%Note that here we consider acceptable encodings (rather than weak acceptable encodings) as defined in Definition
%\ref{def:resenc}.

\begin{theorem}
Let ${\cal G}$ be a class of goals such that if  $H$  is a head of a rule then $K \in {\cal G}$ for any $K\subseteq H$. Then, for n$\geq 2$, there exists no acceptable encoding for qualified answers of CHR$_{n,t}$ into CHR$_1$ for the class ${\cal G}$.
\end{theorem}

\begin{proof}
The proof will proceed by contradiction. Assume that there exists an acceptable encoding for qualified answers
$\encoding{\ } :{\cal P}_{n,t} \rightarrow {\cal P}_{1}$ and $\encoding{\ }_g:{\cal G}_{n,t} \rightarrow {\cal G}_{1}$
 of CHR$_{n,t}$ into CHR$_1$ for the class of goals ${\cal G}$ and let $P$ be the program consisting of the single rule:
$$\simp{r}{H, H}{{\tt true}}{k}$$ where $k$ is an atomic user-defined constraint. The goal $(H, H)$ in $P$ has a qualified answer $k$ (note that for each goal $G$, $P$ has no trivial data sufficient answers different from $G$).

Therefore, by definition of acceptable encoding for qualified answers, the goal
  $\encoding{(H, H)}_g$ in $\encoding{P}$ has a qualified answer $k$ (with the built-in constraint ${\tt true}$).
Since the compositionality hypothesis implies that  $\encoding{(H, H)}_g$ =
$\encoding{H}_g, \encoding{H}_g$, from Lem\-ma \ref{lem:single} it follows that $\encoding{(H, H)}_g$ in program $\encoding{P}$ has also a qualified answer $(k, k)$, but this answer cannot be obtained in the program with multiple heads thus contradicting one of the hypothesis on the acceptable encoding for qualified answers.
Therefore such an encoding cannot exist.
\end{proof}

From previous theorem and Theorem \ref{th:general} follows that, in general, no acceptable encoding for qualified answers of CHR in CHR$_1$ exists.

\begin{corollary}\label{ch:general1}
Let ${\cal G}$ be a class of goals such that if  $H$  is a head of a rule then $K \in {\cal G}$ for any $K\subseteq H$. Then there exists no acceptable encoding (for qualified answers) of
CHR in CHR$_1$ for the class ${\cal G}$.
\end{corollary}

\section{A hierarchy of languages}\label{chrsec:2vsn}	
	%As mentioned in the introduction, we define CHR$_n$ as the CHR language where rules have at most $n$ heads. In this section we aim to extend the results in \cite{DGM08} by studying the existence of acceptable encodings of CHR$_n$ into CHR$_m$ for $n<m$. We will show that CHR$_n$ generates a hierarchy of languages of increasing expressiveness.

%\subsection{A generalization} \label{sec:nteste}

After having shown that multiple heads increase the expressive power with respect to the case of single heads, it is natural to ask whether considering a different number of atoms in the heads makes any difference. In this section we show that this is indeed the case, since we prove that,  for any $n>1$,  there exists no acceptable encoding (for qualified answers) of CHR$_{n+1}$ into
CHR$_n$. Thus, depending on the number of atoms in the heads, we obtain a chain of languages with increasing expressive power.

In order to obtain this generalization,  we need to streng\-then the requirement on acceptable encodings  --- only for data sufficient answers --- given in Definition \ref{def:resenc}. More precisely, we now require that goals are unchanged in the translation process. This accounts for a ``black box'' use of the  program: we do not impose any restriction on the program encoding, provided that the interface remains unchanged.
Hence, in the following theorem we call ``goal-preserving acceptable  encoding'' an
acceptable encoding (according to Definition \ref{def:resenc}) where the function $\encoding{G}_g $ which translates goals  is the identity.

%%\revcinzia{Aggiunta definizione formale come richiesto da rev3}
%\revmau{Tolta definizione formale e aggiunto riga di commento}

%\begin{definition}[Acceptable encoding with identity]\label{def:resencrestr}
%Let ${\cal G}$ be  a class of CHR goals. An {\em acceptable encoding} (of CHR into CHR$_1$, for the class of goals ${\cal G}$) is a pair of mappings $\encoding{\ } :{\cal P}_{CHR} \rightarrow {\cal P}_{CHR_1}$ and $\encoding{\ }_g:{\cal G}_{CHR} \rightarrow {\cal G}_{CHR_1}$ which satisfy the following conditions:
%\begin{itemize}
% \item ${\cal P}_{CHR}$ and ${\cal P}_{CHR_1}$  share the same CT;
% \item for any goal $(A,B) \in {\cal G}_{CHR}$, $\encoding{A,B}_g = A,B$ holds. We also assume that the built-ins present in the goal are left unchanged;
% \item Data sufficient answers are preserved for the class of goals ${\cal G}$, that is, for all $G\in {\cal G}$, %\subseteq {\cal G}_{CHR}$,
% ${\cal SA}_{P}(G) =  {\cal SA}_{\encoding{P}}(\encoding{G}_g)$ holds.
%\end{itemize}
%\end{definition}

We have, then, the following result where we use the notation of Definition \ref{def:dialects}.

%\begin{theorem*}
% Let ${\cal G}$ be the class of all possible goals. When considering data sufficient answers there exists no  acceptable  encoding with identity of CHR$_{n+1}$ in CHR$_n$ for the class ${\cal G}$.
%\end{theorem*}

\begin{theorem}
 Let ${\cal G}$ be the class of all possible goals.
 There exists no goal-preserving acceptable encoding of CHR$_{n+1,d}$ in CHR$_n$ for the class ${\cal G}$.
\end{theorem}

\begin{proof}
The proof will proceed by contradiction. Assume that there exists a goal-preserving acceptable encoding of CHR$_{n+1,d}$ in CHR$_n$ for the class ${\cal G}$ and let $P$ be the following CHR$_{n+1,d}$ program:
$$\simp{rule}{h_1 \dots h_{n+1}}{{\tt true}}{d}$$
where $V=Fv(h_1 \dots h_{n+1})$, $d$ is a built-in  constraint different from ${\tt false}$ (i.e. $CT \models d \not \leftrightarrow {\tt false}$ holds) such that $Fv(d)$ $\subseteq V$. Hence given the goal $G = h_1 \dots h_{n+1}$ the program $P$ has the data sufficient answer $d$.

Observe that every goal with at most $n$ user defined constraints has no data sufficient answer in $P$.
Now consider a run of $G$ in $\encoding{P}$ (where $\encoding{P}$ is the encoding of the program $P$) with final configuration $\tuple{\emptyset, \emptyset, d'}$, where $CT \models\exists_{-V} (d') \leftrightarrow d$:
$$\delta=\tuple{G, \emptyset, \emptyset} \rightarrow^{*}  \tuple{H_i, G_i, d_i}  \rightarrow \tuple{H_{i+1}, G_{i+1}, d_{i+1}}
\rightarrow^{*} \tuple{\emptyset, \emptyset, d'} \nar, $$
where, without loss of generality, we can assume that in the derivation $\delta$, for any configuration $\tuple{H', G',c'}$ we can use either a Simplify or a Propagate transition only if $H'$ does not contain built-ins and $G_i$ is the last goal to be reduced in the run by using either a Simplify or a Propagate transition. Therefore $G_i$ has at most $n$ user-defined constraints, $H_i= \emptyset$  and let $r \in \encoding{P}$ be the last rule used in $\delta$ (to reduce $G_i$).
Since $d$ is a built-in constraint, $r$ can be only of the following form  $H \Leftrightarrow C \mid C'$, where $H$ has at most $n$ user defined constraints. In this case $G_{i+1}=\emptyset$ and $H_{i+1}$ contains only built-in predicates.
 Then $$CT \models d_i \rightarrow \exists_{Fv(H) }(( G_i=H )\wedge C) \mbox{ and }$$
$$CT \models (d_i\wedge C' \wedge( G_i=H)) \not\leftrightarrow {\tt false}.$$

By construction the goal $(G_i, d_i)$ has the data sufficient
$\exists_ {-Fv(G_i, d_i)} (d')$ in $\encoding{P}$. But the goal $(G_i, d_i)$ has no data sufficient answer in $P$
thus contradicting one of the hypothesis on the goal-preserving acceptable  encoding. Therefore such an encoding cannot exist. \end{proof}

Similarly to the development in the previous section, we now consider the case where the program has only qualified answers and no trivial data sufficient answers. Notice that in this case we do not require anymore that the translation of goals is the identity (we only require that it is compositional, as usual).

\begin{theorem}\label{the:nqualified2}
Let ${\cal G}$ be a class of goals such that if  $H$  is a head of a rule then $K \in {\cal G}$ for any $K\subseteq H$. There exists no  acceptable  encoding for qualified answers of CHR$_{n+1,t}$ in CHR$_n$ for the class ${\cal G}$.
\end{theorem}
\begin{proof}
 The proof is by contradiction. Assume that there exists an acceptable encoding for qualified answers
$\encoding{\ } :{\cal P}_{n+1,t} \rightarrow {\cal P}_{n}$ and $\encoding{\ }_g:{\cal G}_{n+1,t} \rightarrow {\cal G}_{n}$ of CHR$_{n+1,t}$ in CHR$_n$  for the class of goals ${\cal G}$ and let $P$ be  the following CHR$_{n+1,t}$ program:
$$\simp{rule}{h_1 \dots h_{n+1}}{true}{k}$$
where $V=Fv(h_1 \dots h_{n+1})$ and $k$ is an atomic user defined constraint such that $Fv(k)$ $\subseteq V$.
Hence given the goal $G = h_1 \dots h_{n+1}$ the program $P$ has only the qualified answer $k$ and since $k$ is an atomic user defined constraint, we have that $k \neq (h_1 \dots h_{n+1})$.

Observe that every goal with at most $n$ user defined constraints has only itself as qualified answer in $P$.
Then since the encoded program has to preserve all the qualified answers in the original $P$, every goal $\encoding{G_n}_g$, where $G_n$ has at most $n$ user defined constraints, has a qualified answer $G_n$ in $\encoding{P}$.

Therefore, if we denote by  $G_n = h_1 \dots h_{n} $, by previous observation and by definition of qualified answers, we have that there exist two derivations
$$\la \encoding{G_n}_g , \emptyset, \emptyset \ra \rightarrow^{*}\la \emptyset, G'_n, d \ra  \nar \mbox{ and }
\la\encoding{h_{n+1}}_g , \emptyset, \emptyset \ra\rightarrow^{*} \la \emptyset, h'_{n+1} , d' \ra \nar,$$
such that $$CT \models G_n
\leftrightarrow \exists _{-Fv(\encoding{G_n}_g)} (G'_n \wedge  d)\mbox{ and } CT \models h_{n+1} \leftrightarrow \exists _{-Fv(\encoding{h_{n+1}}_g)} (h'_{n+1} \wedge d').$$

Without loss of generality, we can assume that
\[Fv(G'_n, d) \cap Fv(h'_{n+1} , d') \subseteq Fv(\encoding{G_n}_g) \cap Fv(\encoding{h_{n+1}}_g).\]

Now consider the goal $G$, from what previously said we have that:
$$\la \encoding{G}_g , \emptyset, \emptyset \ra \rightarrow ^{*} \la \encoding{h_{n+1}}_g, G'_n, d \ra$$
but we also know that $\la\encoding{h_{n+1}}_g , \emptyset, \emptyset \ra\rightarrow^{*} \la \emptyset, h'_{n+1} , d' \ra \nar$ and this  cannot be prevented by any step in the  previous run, thus we obtain:
$$\la \encoding{G}_g , \emptyset, \emptyset \ra \rightarrow ^{*}  \la \emptyset, (G'_n, h'_{n+1}), d \wedge d'\ra,$$ where
$CT \models  G
\leftrightarrow \exists _{-Fv(\encoding{G}_g)} (G'_n \wedge  h'_{n+1}\wedge  d \wedge d')$.
Since $G$ is not a qualified answer for the goal $G$ in $P$ and since $\encoding{P}$ is an acceptable encoding of $P$ in CHR$_n$, we have that there exists $\{h'_{j_1}, \ldots h'_{j_s} \}\subseteq \{G'_n, h'_{n+1}\}$, with $s \leq n$,  such that
$\la \emptyset, (h'_{j_1}, \ldots h'_{j_s} ), d \wedge d'\ra \rightarrow  \la G', H', d'' \ra$ in $\encoding{P}$.
Then, since
 $CT \models G \leftrightarrow \exists _{-Fv(\encoding{G}_g)} (G'_n \wedge  h'_{n+1}\wedge  d \wedge d')$, we have that there exists
 $\{h_{j_1}, \ldots h_{j_s} \}\subseteq G$ such that
\[CT \models h_{j_1}, \ldots h_{j_s}
\leftrightarrow  \exists _{-Fv(\encoding{h_{j_1}, \ldots h_{j_s}}_g)}(h'_{j_1}, \ldots h'_{j_s} \wedge d \wedge d')\] and therefore $h_{j_1}, \ldots h_{j_s}$ is not a qualified answer for
$\encoding{h_{j_1}, \ldots h_{j_s}}_g$ in $\encoding{P}$
(since it is always possible to make another derivation step from $h_{j_1}, \ldots h_{j_s}$ in  $\encoding{P}$).

But, by previous observations, the same goal  has itself as answer in $P$ thus contradicting the fact that
there exists an acceptable encoding for qualified answers of CHR$_{n+1,t}$ in CHR$_n$.
\end{proof}

Notice that an immediate generalization of previous Theorem \ref{the:nqualified2} implies that also under the weaker assumption of compositionality (rather than identity) for the translation of goals, no acceptable encoding for qualified answers for general  $CHR_{n+1}$ programs (including programs with data sufficient answers)  into $CHR_n$ exists.
Therefore, from  Theorem~\ref{the:nqualified2} we have immediately the
following.

\begin{corollary}\label{cor:nqualified2}
Let ${\cal G}$ be a class of goals such that if  $H$  is a head of a rule then $K \in {\cal G}$ for any $K\subseteq H$. There exists no  acceptable  encoding for qualified answers of CHR$_{n+1}$ in CHR$_n$ for the class ${\cal G}$.
\end{corollary}

It is also worth noticing that the previous results depend on the class of considered goals. In fact,
if we limit the class of intended goals for a program and assume that some predicates in the
translated program cannot be used in the goals, then 
%It is also worth noticing that for the correctness of previous results it is essential to consider all the possible goals (which can be expressed in the given signature). In fact, if we limit the class of intended goals for a program and assume that some predicates in the translated program cannot be used in the goals, 
one can easily encode a  $CHR_n$ program into a
$CHR_2$ one. Consider for example the program consisting of the single rule
$$\simp{rule}{h_0 \dots h_{n}}{C}{B}$$ and assume that the only valid goal for such a program  is $h_0 \dots h_{n}$, while  $i_1, \dots, i_n$ are fresh user-defined constraints that cannot be used in the goals. Then the following
$CHR_2$ program is equivalent to the original one
\begin{align*}
 \texttt{r$_1$} \ &@ \ h_0, h_1 \Leftrightarrow i_1\\
 \texttt{r$_2$} \ &@ \ h_2, i_1 \Leftrightarrow i_2\\
 \dots\\
 \texttt{r$_n$} \ &@ \ h_n, i_{n-1} \Leftrightarrow C \mid B
\end{align*}

%OLD However, we find that this restriction on fresh user-defined constraints to be used only in the encoding is too strong. In fact, even though  from a pragmatic point of view it is meaningful to define a class of acceptable goals for a program, logic programming languages (and CHR) do not provide any scoping mechanism which imposes visibility restrictions on the names used in a program. Hence, at least from a theoretical point of view, all the (predicate) names used in a program can be considered in the goals. Indeed, essentially all the existing semantics for logic languages take this point of view and define the semantics of a program in a goal independent way, referring to all the possible predicates used in a program (or in the given signature).

This restriction on fresh user-defined constraints to be used only in the encoding is rather strong, since all logic programming languages (including CHR) allow to use in the goals all the predicate names used in the program. In fact, essentially all the existing semantics for logic languages define the semantics of a program in a goal independent way, referring to all the possible predicates used in a program (or in the given signature). Nevertheless, from a pragmatic point of view it is meaningful to define a class of acceptable goals for a program and then to consider encoding, semantics etc, only w.r.t. that class of goals.  In this respect it would be interesting to identify weaker conditions on goals and predicate names which allow to encode $CHR_{n+1}$  into $CHR_{n}$ (see also Section \ref{chrsec:conclusions}).

\section{Conclusions and Related works}\label{chrsec:conclusions}
	
In this paper we have shown that multiple heads augment the expressive power of CHR. Indeed we have seen that the single head CHR language, denoted by CHR$_1$, is not Turing powerful when the underlying signature (for the constraint theory) does not contain function symbols, while this is not the case for CHR.
Moreover, by using a technique based on language encoding, we have shown that CHR is strictly more expressive than CHR$_1$ also when considering a generic constraint theory, under some reasonable assumptions (mainly, compositionality of the translation of goals).
%
%Then we extended this  result  by showing that CHR is more expressive than CHR with single heads also when considering programs which allow qualified answers only. More precisely we have proved that, when considering this class of programs, CHR cannot be encoded into CHR$_1$ assuming that goals are translated in a compositional way and that both the languages use the same theory for built-in constraints.
Finally we have shown that, under some slightly stronger assumptions, in general the number of atoms in the head of rules affects the expressive power of the language. In fact we have proved that CHR$_n$ (the language containing at most $n$ atoms in the heads of rules) cannot be encoded into CHR$_m$, with  $n>m$.

 There exists a very large literature on the expressiveness of concurrent languages, however there are only few papers which consider the expressive power of CHR.  A recent one is \cite{Sn08}, where Sneyers shows that several subclasses of CHR are still Turing-complete, while single-headed CHR without
host language and propositional abstract CHR are not Turing-complete. Moreover
\cite{Sn08} proves essentially the same result given in Theorem
\ref{th:turing2} by using Turing machines rather than Minsky machines. Both Theorems \ref{th:turing} and \ref{th:turing2} were contained in the short version of this paper \cite{DGM08}, submitted before \cite{Sn08} was published and both these results, including the encoding of the Minsky machine, were suggested by Jon Sneyers in the review of an older version  (\cite{DGM08corr}) of \cite{DGM08}. It is worth noting that very similar encoding exists in the field of process algebras. For example, in \cite{BGZ04} an encoding of Minsky machines in a dialect of CCS  is provided which represents the value $n$ of a register by using a corresponding number of parallel processes connected in a suitable way. This is similar to the idea exploited in Section \ref{chrsec:turing}, where we encoded the value $n$ of a registers by using  a conjunction (the CHR analogous of CCS parallel operator) of $n$ atomic formulas.

Another related study  is \cite{SnScDe05}, where the authors show that it is possible to implement any algorithm in CHR in an efficient way, i.e. with
the best known time and space complexity.  This result is obtained by  introducing a new model of computation, called the CHR machine, and comparing it with the well-known Turing machine and RAM machine models. Earlier works by Fr\"uhwirth \cite{Fr01,Fr02} studied the time complexity of simplification rules for naive implementations of CHR.
In this approach an upper bound on the derivation length,
combined with a worst-case estimate of (the number and cost of) rule application attempts, allows to obtain an upper bound of the time complexity. The aim of  all these works is clearly completely different from ours, even though it would be interesting to compare CHR and CHR$_1$ in terms of complexity.

When moving to other languages, somehow related to our paper is  the work by Zavattaro \cite{Zavattaro98} where the coordination languages Gamma \cite{BM93} and Linda \cite{GC92} are compared in terms of expressive power. Since Gamma allows multiset rewriting it reminds CHR multiple head rules, however the results of \cite{Zavattaro98} are rather different from ours, since a process algebraic view of Gamma and Linda is considered where the actions of processes are atomic and do not contain variables. On the other hand, our results depend directly on the presence of  logic variables in the CHR model of computation.
Relevant for our approach is also \cite{BoePal94} which introduces the original approach to language comparison based on encoding, even though in this paper rather different languages with different properties are considered.

In \cite{LV08} Laneve  and Vitale show that a language for modeling molecular biology, called $\kappa$-calculus, is more expressive than a restricted version of the calculus, called nano-$\kappa$,  which is obtained by restricting to ``binary reactants'' only (that is, by allowing at most two process terms in the left hand side of rules, while $n$ terms are allowed in $\kappa$). This result is obtained by showing that, under some specific assumptions, a particular (self-assembling) protocol cannot be expressed in nano-$\kappa$, thus following a general technique
 which allows to obtain separation results by showing that (under some specific hypothesis) a problem can be solved in a language and not in another one (see also \cite{Pal03} and  \cite{VPP07}).
 This technique is rather different from the one we used, moreover also the assumptions on the translation used in \cite{LV08} are different from ours. Nevertheless, since $\kappa$ (and nano-$\kappa$) can be easily translated in CHR, it would be interesting to see whether some results can be exported from a language to another. We left this as future work.

% We are also planning to investigate what happens when the source and the target CHR languages have different theories for the built-ins: we believe that some results hold also in this more general case, however some technical details need to be spelt out.

%It is worth mentioning that in process calculi often separation results are obtained by showing that (under some reasonable hypothesis) a problem can be solved in a language and not in another one.  For example, in \cite{Pal03} the author proves that there exists no \emph{reasonable} encoding from the $\pi$-calculus to the asynchronous $\pi$-calculus by showing that the symmetric leader election problem has no solution in the asynchronous version of the $\pi$-calculus (for a recent overview on the subject see \cite{VPP07}).
%Even though the $\pi$-calculus and the languages considered in  \cite{VPP07}  are rather different from CHR, we believe that we could adapt this approach to our case: in fact we believe that the leader election problem can be implemented by using CHR programs, while it cannot be solved by using CHR$_1$.

We also plan to investigate what happens when considering the translation of CHR
%into real Prolog systems. In fact, (Constraint) Logic programming and Prolog are programming languages quite different from CHR, mainly because they are sequential ones, without any guard mechanism and commit operator. Nevertheless,
since many CHR implementations are built on top of a Prolog system, by using a compiler which translates CHR programs to Prolog. Our technical lemmata about CHR$_1$ can be adapted to what is called \cite{Ap97} ``pure Prolog'',  that is, a logic programming language which uses the leftmost selection rule and the depth-first search. Hence it is easy to show that, under our assumptions, CHR cannot be encoded in pure Prolog. However, implemented ``real'' Prolog systems are extensions of pure Prolog obtained by considering specific built-ins for arithmetic and control, and when considering these built-ins some of the properties we have used do not hold anymore (for example, this is the case of Lemma
\ref{lem:preservarisposte}). Hence it would be interesting to see under which conditions CHR can be encoded in real Prolog systems, that is, which features of real Prolog (which are not present in pure Prolog) are needed to obtain an acceptable encoding of CHR.  Finally we plan to extend our results  to consider specific constraint theories (e.g. with only monadic predicates) and also taking into account the refined semantics defined in \cite{DGS04}. This latter semantics requires further work, because it allows an improved control on computations and some properties that we used do not hold anymore in this case. 
 
\begin{ack}
We thank the reviewers for their precise and helpful comments.  
\end{ack}

\bibliographystyle{acmsmall}
\bibliography{biblio}

 \received
  {April 2009}
  {November 2010}
  {December 2010}

\end{document}